\input harvmac
\newcount\figno
\figno=0
\def\fig#1#2#3{
\par\begingroup\parindent=0pt\leftskip=1cm\rightskip=1cm
\parindent=0pt
\baselineskip=11pt
\global\advance\figno by 1
\midinsert
\epsfxsize=#3
\centerline{\epsfbox{#2}}
\vskip 12pt
{\bf Fig. \the\figno:} #1\par
\endinsert\endgroup\par
}
\def\figlabel#1{\xdef#1{\the\figno}}
\def\encadremath#1{\vbox{\hrule\hbox{\vrule\kern8pt\vbox{\kern8pt
\hbox{$\displaystyle #1$}\kern8pt}
\kern8pt\vrule}\hrule}}

\overfullrule=0pt

\Title{\vbox{\baselineskip12pt
\hbox{hep-th/  }
\hbox{TIFR-TH/01-16}
}}
{\vbox{\centerline{Noncommutative X-Y model and}
\centerline{Kosterlitz Thouless transition }}}
\smallskip
\centerline{Bahniman Ghosh~
\foot{email address : bghosh@theory.tifr.res.in}}
\smallskip
\centerline{{\it Tata Institute of Fundamental Research}}
\centerline{\it Homi Bhabha Road, Mumbai 400 005, INDIA}
\bigskip

\medskip

\noindent
Matrix models have been shown to be equivalent to noncommutative field
theories. In this work we study noncommutative X-Y model and try to
understand Kosterlitz Thouless transition in it by analysing the
equivalent matrix model. We consider the cases of a finite lattice and
infinite lattice separately. We show that the critical value of the
matrix model coupling is identical for the finite and infinite
lattice cases. However, the critical values of the coupling of the
continuum field theory, in the large $ N $ limit, is finite in the
infinite lattice case and zero in the case of finite lattice.

\Date{April, 2001}

\newsec{Introduction}

The BFSS \ref\banks{T. Banks, W. Fischler, S.H. Shenker and L. Susskind,
Phys. Rev. D55 ( 1997 ) 5112, hep-th/9610043. } and IKKT \ref\Iso{N. Ishibashi, H. Kawai, Y. Kitazawa and A. Tsuchiya,
Nucl. Phys. B498 (1997 ) 467, hep-th/9612115.} 
theories have generated recent interest in matrix
models as tools for studying nonperturbative aspects of M-theory and type
IIB superstring theory. In the BFSS formalism, M-theory formulated in
the infinite momentum frame has been conjectured to be equivalent to
the $ N \rightarrow \infty $ limit of a $ U(N) $ supersymmetric
quantum mechanics written as a matrix model with $ N \times N $
hermitian matrices. In the IKKT theory, type IIB Green Schwarz
Superstring action in the Schild gauge has been related to a matrix
model which has been derived as an effective theory of large $ N $
reduced model of  10 - dimensional Super-Yang Mills theory.
In addition, there has been a recent surge of interest in
noncommutative gauge theories after it has been realised that they appear as
limits of D-brane open string theories in the presence of nonvanishing
NSNS B fields \ref\douglashull{A. Connes, M. Douglas and A. Schwarz,
JHEP 9802 (1998) 003, hep-th/9711162; M. Douglas and C. Hull, JHEP 02
(1998) 008, hep-th/9711165, M. Douglas, hep-th/9901146.} ,
\ref\noncommutative{Y.K.E. Cheung and M. Krogh,
Nucl. Phys. B528 (1998) 185, hep-th/9803031; F. Ardalan, H. Arfaei and
M.M. Sheikh-Jabbari, JHEP 02:016 (1999), hep-th/9810072; C.S. Chu and P.M. Ho,
Nucl. Phys. B550 (1999) 151, hep-th/9812219; M.M. Sheikh-Jabbari, Phys. Lett. B 455 (1999) 129, hep-th/9901080 and JHEP 06 (1999) 015, hep-th/9903205; A. Hashimoto and N. Itzhaki, Phys. Lett. B 465 : 142 - 147, 1999, hep-th/9907166; D. Bigatti and L. Susskind, Phys. Rev. D 62 : 066004, 2000,
hep-th/9908056; J. Maldacena and J. Russo, JHEP 9909 : 025, 1999, hep-th/9908134;  N. Seiberg and E. Witten, JHEP 09 (1999) 032, hep-th/9908142. }.   

$ U(N) $ gauge theories in commutative spacetime can be studied, in the t'Hooft large $ N $ limit, by
equivalent $ U(N) $ matrix models \ref\eguchikawai{T. Eguchi and
H. Kawai, Phys. Rev. Lett. 48 (1982) 1063. }    where spacetime dependent fields are 
mapped to spacetime independent $ U(N) $ matrices to obtain a reduced
model from the gauge theory and equality is obtained at the level of
correlation functions of the field theory and the matrix model. The
property of factorization of correlation functions of the gauge theory
in the  $ N \rightarrow \infty $ limit, then guarantees equivalence of
the gauge theory and the matrix model, provided the matrix model is
either quenched (QEK model) or twisted (TEK model) \ref\wquenched{G. Parisi, Phys. Lett. 112B (1982) 463; G.
Bhanot, U. M. Heller, and H. Neuberger, Phys. Lett. 113B ( 1982 ) 47;  S. Das and S. Wadia,
Phys. Lett. 117B (1982) 228;  D. Gross and
Y. Kitazawa, Nucl. Phys. B206 (1982) 440; A. Gonzalez-Arroyo and M. Okawa, Phys. Lett. B120 (1983) 174; T. Eguchi and R. Nakayama, Phys. lett. B122 ( 1983 ) 59; A. Gonzalez-Arroyo and C.P. Korthals Altes, Phys. Lett. B131 ( 1983 )
396 ; A. Gonzalez and M. Okawa, Phys. Rev. D27 ( 1983 ) 2397. }. In the usual
't Hooft limit of the TEK
model, a $ U(N) $ matrix model is equivalent to a $ U(N) $ field
theory (in commutative spacetime)  in a periodic box. In some recent papers \ref\kitazawa{H. Aoki,
N. Ishibashi, S. Iso, H. Kawai, Y. Kitazawa and T. Tada,
Nucl. Phys. B565 ( 2000 ) 176, hep-th/9908141; N. Ishibashi, S. Iso,
H. Kawai and Y. Kitazawa, Nucl. Phys. B573 : 573 - 593, 2000, hep-th/9910004. }, however, it has been shown that
there exists a limit of the twisted reduced models when they
describe quantum field theory on noncommutative spacetime, in particular, a
$ U(mn) $ twisted reduced model is equivalent to a $ U(m)_{\theta} $ gauge
theory (where $ \theta $ is the noncommutativity parameter) on noncommutative spacetime in the limit of $ n \rightarrow \infty $. This limit is, clearly,
different from the t'Hooft large $ N $ limit ( in particular, nonplanar
diagrams survive in this limit while they do not in the t'Hooft limit
) and, in a sense, it converts
part of the noncommutativity of the matrix model into noncommutativity
of spacetime coordinates, which are generated from the spacetime independent
matrices in the reduced model by expanding a general matrix around a
particular classical solution which exists only in the limit $ n
\rightarrow \infty $. Some other aspects of the noncommutative field
theory and matrix
model correspondence have beeen considered in \ref\morita{ A. Schwarz,
Nucl. Phys. B534 (1998) 720, hep-th/9805034.}.  

In this paper we use the relationship between noncommutative field
theory and twisted reduced models to study Kosterlitz Thouless
transition in noncommutative XY model. In the commutative XY model,
realised as a two dimensional system of interacting spins, there exist
vortex configurations in which the order parameter is an integer
multiple of the polar angle with the corresponding integer
being the winding number of the vortex. Such vortex configurations
are, however, not energetically favourable at all temperatures because
of a competition between the energy required to form the vortex and
the entropy of the vortex. Kosterlitz Thouless (KT) transiton
\ref\kostelitz{J.M. Kosterlitz and D.J. Thouless, J. Phys. C 6,
(1973) 1181. } refers to the
transition from a phase of no vortex to a phase of vortices and the temperature at which this transition occurs is called
the KT transition temperature. This transition is characterstic of two spatial dimensions,
since, in this case,
both the vortex energy and the entropy of a single vortex scale as the
logarithm of the system size. In
noncommutative space one can, formally, construct the action for an XY model
that reduces, in the appropriate limit, to the action for commutative
XY model. Going to the reduced model one can, then, show that there
exist vortex configurations for which the energy and entropy in two
spatial dimensions scale as
the logarithm of the system size. One can, by a calculation of
the free energy 
similar to the commutative case, find an expression for the KT
transition temperature, which turns out to be independent of
the noncommutativity parameter $ \theta $.  A brief comment on the
nature of divergence of the noncommutative vortex action is,
probably, in order here. The action scales as the sum of two terms
one linear and the other a logarithm of the system size. By a suitable
regularization, one can extract the logarithm piece for the action
which gives the desired behaviour for the vortex energy.

We organise the paper as follows. In section 2 we develop
our notation and review the relation between noncommutative gauge
theories and twisted reduced models. In section 3 we discuss the
noncommutative XY model and map it to the equivalent matrix model. In
section 4 we find the vortex solution to the XY model and
discuss Kosterlitz Thouless transition in the continuum limit
of the model with infinite lattice size. In section 5 we study the
same for the case of a finite lattice. Section 6 ends with some
concluding remarks.

\newsec{Description of noncommutative field theory in terms of
reduced model}
 
Field theories in noncommutative spaces can be studied by twisted 
reduced models obtained by a 
mapping of the fields on the noncommutative space to 
matrix valued operators. The usual matrix multiplication involving
these operators then defines the way in which fields are multiplied in
the corresponding noncommutative space. Integration of fields over
noncommutative coordinates is then equivalent to tracing over the
corresponding matrix operators.
On a noncommutative space of dimension $ D $, the local coordinates $ x^
{\mu} $ are replaced by hermitian operators  $ X^{\mu} $ obeying the
commutation relations
\eqn\bone{ [{X}^{\mu}, {X}^{\nu}] = i {\theta ^{\mu \nu}}}
where $ \theta ^{\mu \nu } $ are dimensionful real valied c-numbers. 
Fields are defined as functions on this coordinate space $ x^{\mu} $. The
reduced model prescription is equivalent to finding matrix
representation of the coordinate space, that respects the basic
commutation relation {\bone } , and of fields on this space. However, it turns
out, that {\bone } does not have finite dimensional representations.
 For the purpose of doing explicit calculations one
puts the field theory of interest on a lattice with some lattice
spacing. Noncommutativity then implies that the lattice is
periodic \ref\Itzhak{ I.Bars, D. Minic, Phys. Rev. D 62 : 105018, 2000, hep-th/9910091. }, \ref\Ambjorn{ J. Ambjorn,
Y. M. Makeenko, J. Nishimura and R. J. Szabo, JHEP 9911 (1999) 029,
hep-th/9911041. }, \ref\mak{J. Ambjorn,
Y. M. Makeenko, J. Nishimura and R. J. Szabo,  Phys. Lett. B 480 (2000) 399, hep-th/0002158 and JHEP 0005 : 023, 2000, hep-th/0004147. } .
   
 As an example, let us consider a scalar
field defined on the noncommuting space $ X^{\mu} $ with
lattice spacing $ a $ and linear size $ L $. We will specialize to two
dimensions and hence $ \mu = 1,2. $ We impose periodic boundary
condition on the lattice and construct, formally, the matrices 
 $ \Gamma _{1} \equiv e^{iaB X^ {2}} , \Gamma _{2} \equiv e^{iaB X^
{1}}  ( $ where $ B = - { 1 \over {\theta ^{12}}} $ ),  which generate translations by one lattice
spacing in the directions $ X^1 $ and $ X^2 $ respectively. In a periodic lattice it is possible to obtain finite dimensional representation of the $ \Gamma _ {\mu} $ s even though the $ X^{\mu} $ s themselves are infinite dimensional. We consider an $ N $ dimensional representation of the $ \Gamma _{\mu} $ s which satisfies the following relation,  
\eqn\bfour{ \Gamma _1 \Gamma _2 = e^ {i \alpha } \Gamma _2 \Gamma _1}
\eqn\bfourone{ \Gamma _{\mu}^{N} = 1,~~~~~~~~~  (\mu = 1,2) } 
where $ \alpha = { {2 \pi} \over N} $. For an explicit construction of
these matrices we refer to \ref\thooft{ G. 't Hooft, Commun. Math. Phys., 1981, 267. }.
All operators in this space can be written as linear combinations of integral
powers of these matrices (since the set of all translations
generate the full quantum space)        
\eqn\bfive{ {\bar {\phi}} = \sum_{n_x, n_y = 0}^{N-1}  \phi_{n_x n_y}
 {\Gamma_1^{n_y} } {\Gamma_2^{n_x} }  e^{-{i \over 2 } n_x n_y \alpha }
\equiv \sum_{k_x k_y} {\tilde \phi }(k_x , k_y) e^{- i k_{\mu} X^{\mu}}}
where $ \phi_{n_x n_y} = {\tilde \phi}(k_x , k_y) $ and the $ k $ s belong to the discrete ( because the lattice size is
finite) dual lattice.
The corresponding field in noncommutative space is then defined as a
fourier tranform of the scalar valued coefficient $ {\tilde \phi} (k) $ 
\eqn\bthree{ \phi (x, y) = \sum_{k_x  k_y} {\tilde {\phi }}(k_x ,  k_y) e^{-ik_{\mu} x^{\mu} }}
This gives a new coordinate space $ x^{\mu} $  which one interprets as the
semiclassical limit of $ X^{\mu }$. 

We label the lattice, with  $ N_0 $ $ ( L = N_0 a ) $ lattice
sites along each dimension,  by the set of dimensionless numbers $ ({\tilde x},
 {\tilde y}) $ . The dimensional position coordinates on the lattice are then
given by $ x = {\tilde x} a $ , $ y = {\tilde y } a $.  The
finite lattice size implies that the dimensional momenta $ k_x $ and $
k_y $ are quantized in inverse units of the lattice size,
\eqn\bsix{ k_x = {{ 2 \pi n_x } \over {N_0 a }},~~~~ k_y = {{ 2 \pi n_y }
\over {N_0 a }}}  
where $ n_x $ and $ n_y $ are integers. We choose $ N = N_0 $ and, hence, fourier expand the field $
\phi $ in terms of dimensionless coordinates $ (\tilde x, \tilde y) $
and dimensional coordinates $ (x, y) $  on the
lattice as follows,
\eqn\fifty{ { {\phi}}(\tilde x, \tilde y) = \sum_{n_x ,n_y } { \phi
_{n_{x} n_{y}}} e^{-i\alpha ({n_x} \tilde x + {n_y}\tilde y)} = \phi(x,y) =
\sum_{k_x , k_y } {
{\tilde \phi} (k_x , k_y)} e^{-i(k_{x}x + k_{y}y)}}
where $ \phi _{n_{x} n_{y}} = {\tilde \phi }({k_x}, k_{y}). $ 

  To go to the continuum, we take the
limit of vanishing lattice spacing while keeping the dimensional
noncommutativity parameter $ \theta $ fixed. One can determine how $
\theta $ scales with the lattice spacing $ a $ and the lattice size $
L = N_0 a = Na $ as follows. We consider the product of two matrix model operators
$ \bar \phi_1 $ and $ \bar \phi_2 $ given by
\eqn\fiftyone{\bar \phi_{1} \bar\phi_{2} = \sum_{{n_i}_x ,{n_i}_y = 0 }^{ N-1} {\phi_{{n_1}_x,{n_1}_y}}
{\phi_{{n_2}_x,{n_2}_y} } {\Gamma _1^{{n_1}_y + {n_2}_y}}{\Gamma _2^{{n_1}_x + {n_2}_x}} e^{-{i
\over 2}({n_1}_x + {n_2}_x)({n_1}_y + {n_2}_y) \alpha} e^{{-i \over 2}\alpha ({n_1}_x {n_2}_y
- {n_2}_x {n_1}_y )}}
Recalling the map from the reduced model to the noncommutative field
theory we obtain the following equality
\eqn\fiftytwo{ {\theta^{12}}( {k_{1x}}{k_{2y}} - {k_{1y}}{k_{2x}})  =
\alpha (n_{1x} n_{2y} - n_{1y} n_{2x})}  
from which we obtain the following scaling equation  for $ \theta $ 
 
\eqn\bseven{ \theta^{12} \equiv \theta = {{L a} \over {2 \pi }} = {{ N a^2 } \over { 2
\pi }}}  
where we have used the relation $ L = N a $. {\bseven } is also
verified from { \bone } and { \bfour }.      
The continuum limit is taken such that $ a \rightarrow
0. $  Then, in order to keep $ \theta $ fixed,  one has to take $ N
\rightarrow \infty $ with $ L a = N a^2 $ fixed. Thus, in the present
case ( in section 5 we will take a different continuum limit such that
the dimensional lattice size is fixed ), the
continuum limit forces one to go to the limit of infinite lattice
size . In the continuum limit, the map between the noncommutative field $
\phi $ and the matrix model operator $ \bar \phi $ remains the same as
in {\bfive }  and {\bthree } with the sum over the discrete set $ k $ being
changed to integration over continuous values of $ k $. The derivative
of the field $ \phi $ is then mapped to the following commutator in
the reduced model  
\eqn\bnine{ i { \del _{ \mu  }} \phi  \leftrightarrow [P_{\mu }, \phi ]} 
where $ P_{\mu} = {(\theta ^ {-1})}_{\mu \nu } {X }^{\nu } \equiv
{ B }_{\mu \nu } {X }^{\nu } $ and vector indices are raised or lowered
with the metric $ g_{\mu \nu } = \delta _ {\mu \nu } $ .
    
Finally, we mention that the following relation holds between the
integrals of fields on the noncommutative space and traces of
operators in the reduced model,
\eqn\beight{Tr {{\bar \phi}} = N {\phi_{00}} = { 1 \over N} \sum_{\tilde x
\tilde y} \phi (\tilde x, \tilde y)  = { 1 \over {2 \pi \theta } } {\int
 dx dy \phi (x , y )}}

\newsec{Non commutative XY model}

In this section we will consider the noncommutative XY model and try to
understand the nature of Kosterlitz - Thouless
transiton in it. The continuum action for the model 
is given by
\eqn\aone{ S = {1 \over {g_{NC}^2}} \int d^2 x ( \del _{\mu} U^+ \del
^{\mu} U )}
where
\eqn\atwo{ U^{*} (x) * U(x) = 1 }
Here $ U $ ( $ U^* $ is the complex conjugate of $ U $ ) is a scalar field on the two dimensional noncommutative space
$ ( x^1, x^2 ) $  and product of fields implies the star product. To see
that this model is indeed the noncommutative analogue of the XY model
one first uses \atwo\  to prove that  $ U(x) $ can be written as
$ U(x) = ({e^{i \theta (x)}})_{*} $,  where $ \theta (x) $ is a new
field. One can easily check that  the action in \aone\   then takes the form
\eqn\bten{ S = { 1 \over { g^2_{NC}}} \int d^2 x (\del _\mu \theta *
\del ^\mu \theta + {i^2 \over {2!}}( \theta * \del _\mu \theta * \del
^\mu \theta - \del _\mu \theta * \del ^\mu \theta * \theta ) +
O({\theta}^4) + \cdots )  }
In the commutative limit, when star products are replaced by
ordinary multiplication of functions, the action takes the form $ S = { 1
\over { g^2_{NC}}} \int d^2 x (\del _\mu \theta)^2 $ which, one easily
recognizes, is the form for the dimensionless energy in 
the corresponding statistical mechanical XY model with $ \theta (x) $
denoting the order parameter.   

The corresponding reduced model is obtained by replacing derivatives
by commutators, integrals by traces and star products by ordinary
matrix products according to the prescription
given above and  is, therefore,  given by the following action
\eqn\athree{ S = - {1 \over {g^2}} Tr [P_{\mu}, U^+ ] [P^{\mu}, U]}
where 
\eqn\afour{ U^+  U = U U^+ = 1  ~~{\rm and } ~~   {g^2 Na^2} = g_{NC}^2}
We define creation and annihilation operators (we will see later that
B is negative in our convention) as follows 
\eqn\five{ a = { 1 \over {\sqrt {-2B}}} (P_1 + i P_2) ~~ {\rm and }~~
 a^+ =  { 1 \over {\sqrt {-2B}}} (P_1 - i P_2)}
where
\eqn\seven{ [P_1 , P_2] = - i B_{12} \equiv - iB }
so that $ a $ and $ a^+ $ satisfy the usual commutation relation of
the annihilation-creation operators of a one dimensional harmonic oscillator
\eqn\eight{ [a, a^+] = 1 }
In terms of these variables the action takes the following form
\eqn\anine{ S =  {2B \over{g^2}}  Tr ( [a^+ , U^+][a,U] + [a, U^+][a^+, U])}   
and the equation of motion for the field $ U $ is given by
\eqn\aten{[a^+ , U^+ [a, U ]] + [a, U^+ [a^+ , U]] = 0}

\newsec{Vortex Solution in noncommutative XY model }

There exists a solution in this model which behaves as a vortex solution
in the commutative limit. The solution is given by
\eqn\aeleven{ U_0 = a^n {1 \over {\sqrt {a^{+ n} a^n}}}}
This is the vortex solution for the higgs field in the abelian higgs model with an appropriate gauge field \ref\Witten{ E. Witten,
hep-th/0006071; 
A. P. Polychronakos, Phys. Lett. B 495 : 407 - 412, 2000, hep-th/0007043; D.P. Jatkar, G. Mandal,
S.R. Wadia, JHEP 0009:018,2000,hep-th/0007078. }  but also happens to be the vortex solution in our model in the absence of gauge field. 
To see that this is a vortex solution in the commutative limit we
recall the definition of the creation and annihilation
operators $ a $ and $ a^+ $ ,
\eqn\twelve{ a = {\sqrt {- {B \over 2} }}( X^2 - i X^1 )
, ~~~~ a^+ = {\sqrt {-{B \over 2}}}( X ^2
+ i X^1 )  }
Now  $ X ^ {\mu } $ maps to a new coordinate space $ x^ {\mu} $
interpreted as the semiclassical limit of $ X^ {\mu} $. In this
new coordinate space the solution  $ U_0 $ looks like
\eqn\beleven{ U_0(z , \bar z) = {{z ^n} \over {(z \bar z)^ {n \over 2}}}} 
where $  z = (x^2 - i x^1) $ and $ \bar z $ is the complex conjugate
of $ z $.
Writing $ z $ in polar coordinates as $ z = r  \exp (i \phi) $, we regain
the following behaviour of the above solution, namely,
\eqn\thirteen{ U_0(r, \phi) = e ^{in \phi }}
which is a vortex in two dimensions with winding number $  n $. 
In the commutative XY model in two dimensions it can be easily shown,
by a straightforward calculation, that the action for the vortex
behaves as $ ln L $ where $ L $ is the linear dimension of the two
dimensional space . We evaluate the
noncommutative action in the following way. We first consider a
complete set of states of the number operator $ a^+ a $ and denote the
individual states by $ |m> $, with $ m $ running from $ 0 $ to $ \infty
$. We then evaluate the action as a trace over this complete set of
states \ref\trace{D.J. Gross and N. Nekrasov, JHEP 0007 : 034, 2000, hep-th/0005204. } . However, it turns out that a complete analytic behavior of
this action is not possible to obtain because of the infinite sum over
the oscillator states. We regularize by introducing an upper cutoff $
R $ on
the oscillator number $ m $ and evaluate the action as a sum only over
states ranging from $ |0> $ to $ |R> $. This prescription for
regulating  the action can be interpreted as an infrared regulation
in the noncommutative theory because of the following reason. The
spread ( root mean square deviation ), in postion space $ (x^1, x^2) $, of the wavefunction
corresponding to the state $ |m> $ can be calculated to be $ {\sqrt {m
\theta }} $. The spread must be bounded by the system size $ L = \sqrt
{ 2 \pi N \theta } $. Equating the maximum spread with the system size we
obtain $ R = 2 \pi N $ . Since the spread depends on the oscillator number,
imposing an infrared cutoff ( or fixing a particular system size )
implies a system size dependent cutoff on the harmonic oscillator
level number $ m $ and is expressed by the relation $ L = \sqrt { R \theta } $ . Without the cutoff,  we calculate
the action for this vortex solution as follows,
\eqn\fourteen{\eqalign{S & = {2B \over {g^2}} {\rm Tr}[a^\dagger U_0^\dagger a U_0
+a^\dagger U_0 a U_0^\dagger - (aa^\dagger + a^\dagger a)]\cr
&= {2B \over {g^2}}[\sum_{m=n}^\infty{\sqrt{m(m-n)}}
+\sum_{m=0}^\infty{\sqrt{m(m+n)}} - \sum_{m=0}^\infty(2m+1)]\cr
&= {2B \over {g^2}}[2\sum_{m=0}^\infty{\sqrt{m(m+n)}} - \sum_{m=0}^\infty(2m+1)]}}
Here the limits on the first sum in the second line comes because
\eqn\fifteen{\sum_{m=0}^\infty <m|a^\dagger U_0^\dagger a U_0|m>
= \sum_{m=n}^\infty{\sqrt{m(m-n)}}}
and we have shifted the summed over integer to go from the second
to the third line and used the following relations,
\eqn\btwelve{ a|m> = {\sqrt m }|m-1>~, ~~~a^+|m> = {\sqrt {m+1}}|m+1>} 
\eqn\bthirteen{ U_0 |m> = | m - n>~, ~~~U_0^+ |m> = |m + n >}

 Since we are interested in the way the action
scales with the system size or, in other words its large distance
behaviour, we shall try to investigate the behaviour, for large
oscillator numbers, of the sum over oscillator levels in the action.    

Expanding the first term in the action as
\eqn\sixteen{{\sqrt{m(m+n)}}= m(1 + \half {n\over m} - {1\over
8}{n^2\over m^2}
+ \cdots)}
we get
\eqn\seventeen{S = {2B \over {g^2}} \sum_{m=1}^\infty[(n-1) - {1\over 4}{n^2\over m}
+\cdots]}
The first term is linearly divergent            and the second term
is     logarithmically divergent.

One way to make sense of these sums is to use zeta function regularization,
where
\eqn\eighteen{ \zeta (s) = \sum_{n=1}^\infty {1\over n^s}}
While this is divergent for $s=0$ the analytic continuation of
$\zeta(z)$
for arbitrary complex $z$ is finite and equal to $-\half$. Therefore,
it is possible to assign a finite value to the 
first
term. The second term is genuinely divergent and is in
fact
log divergent. Introducing the cutoff $ R $ on oscillator level $ m $
at this stage shows that the action diverges as $ \ln R = \ln {L^2
\over \theta } $. Thus we recover the usual behaviour that the
action (or the dimensionless energy in the statistical mechanical viewpoint)
for a vortex scales as the logarithm of the system size.    

One can follow a more rigorous method for finding  the asymptotic expansion
in $ R=2 \pi N $  of the action by using the Euler - MacLaurin formula,
\eqn\sixtyone{ \sum_{m = 0}^{R} f(m)  = {1 \over 2 } f(R) + {\int
_{0}^{R} dt f(t)} + C + [ {B_2 \over {2!}}f^{(1)}(R) - {B_3 \over
{3!}}f^{(2)}(R) + {B_4 \over {4!}}f^{(3)}(R) - \cdots ]}
where  $ C $ is a constant and $ B $ s are Bernoulli numbers.
Using $ f(m) = \sqrt {m(m+n)} $ in the above formula, we get for our action, 
\eqn\sixtytwo{ S = {2B \over {g^2}}[ R(n-1) - { n^2 \over 4 } {\ln R} + O(1/ R ) +
\cdots] } 
This is the same expression as obtained from \seventeen.\


A  better regularization \ref\qana{A.J. Macfarlane, J.Phys. A22 (1989)
4581; L.C. Biedenharn, J.Phys. A22 (1989) L873; E.E. Donets,
A.P. Isaev, C. Sochichiu, M. Tsulaia, JHEP 0012 : 022, 2000, hep-th/0011090.  }  is obtained by using the q analogue of the  harmonic  oscillator
operators $ a $ and $ a^+ $ denoted by $A$ and $A^+$ respectively,
\eqn\nineteen{ A = a { \sqrt { {[M]_q} \over M }}~~,~~~
A^+=a^+{ \sqrt { {[M]_q} \over M }}}
where 
\eqn\twentyone{[M]_q ={{ q^{M} - q^{-M}} \over {q - q^{-1}}}}
where q is $ 2R$-th root of unity and the usual oscillator limit is the one for $ q \rightarrow 1
$ or $ R \rightarrow \infty $. Replacing $ a, a^+ $ by $ A , A^+ $ respectively, and
evaluating the sum over osclliators we get the following
expression for the action,
\eqn\bfifteen{ S = {2B \over {g^2}} \sum_{m=0}^{\infty} (2 {\sqrt {[m][m+n]}} -
[m+1] - [m])}
The infinite sum is automatically cutoff because $ [R]_{q} = 0 $ for $
q^{2R} = 1 $ and, hence, one does not have to put in a cutoff by hand.
The action can, in principle, be evaluated, as a function of the
cutoff $ R $, by numerical methods. 


A small note on the Kosterlitz - Thouless (KT) transtion for this model is, probably, in order here.
As noted earlier the field theory action  of the vortex varies with  $ R $ as
\eqn\btwentyfour{ S = -{{  B n^2} \over {2 g^2}  } ln R = {
{\pi n^2 }\over {g^2 _{NC}}} ln R = {{\pi n^2 }\over {g^2 _{NC}}} ln
{L^2 \over \theta}  }
The apparent negative sign in the first expression for the action may be
surprising. However, we note that with our choice of conventions 
\eqn\twohundred{ \theta ^{\mu \nu} = {{N a^2} \over {2 \pi }} \epsilon
^{\mu \nu},~~~ B^{\mu \nu } \equiv {(\theta ^{-1})^{\mu \nu}}   = - {{2 \pi }\over {Na^2}} \epsilon ^{\mu
\nu}       } 
from where it easily follows that $ B $ is a negative number
\eqn\twohundredthree{ B \equiv B^{12} = - {1 \over {\theta ^{12}}}
\equiv -{1 \over \theta}}
One can, of course, obtain a positive $ B $ by a change of convention
which involves interchanging the roles of   $ X^1 , X^2 $ but since
the action  { \aone } is symmetric in $ x^1 $ and $ x^2 $ it still
remains positive.
The entropy $ \bar S $ of a vortex depends on the number of available
locations for the vortex. Because of the noncommutativity of the
coordinates, a vortex cannot be localized in any region with area
smaller than $ \theta $, since, the product of uncertainties of the two
noncommuting coordinates is of order $ \theta $. This is true because from the relation
\eqn\art{ [X^1 , X^2] = i \theta }
it follows that 
\eqn\ghj{ \Delta X^1 \Delta X^2  \geq {\theta \over 2 }}
where $ \Delta X^i $ denote the root mean square deviations of the
coordinates $ X^i $ taken over any state of the system. 
Therefore, for an area $ A = X^1  X^2 $ of order $ \theta $, the
minimum uncertainty, $ \Delta A $, is of order $ \theta $ which implies that
the centre of a vortex cannot be localized to a region of area less than order
$ \theta $. 
Hence, the number of
positions available to a vortex in a region of size $ L^2 $ is of order $ L^2
\over \theta $, which implies that the entropy should scale as $ \ln
{ L^2 \over \theta }. $  Therefore, the dimensionless Helmholtz free energy
is given, in terms of the couplings $ g $ and $ g_{NC} $   by 
\eqn\twentsix{
F \equiv S -  {\bar S} = - { {  B n^2 }\over { 2 g^2 } }ln {L^2
\over \theta } -  ln {L^2 \over \theta }     =  {{  \pi n^2 }\over { g^2 _{NC}} }ln {L^2
\over \theta } -  ln {L^2 \over \theta } }
Hence, the
couplings $ g_c $ and $ (g_{NC})_c $ at which it is first favourable to produce vortices is
given by
\eqn\twentyfive{  g^2_c = -{{B n^2} \over 2 }, ~~~~  (g _{NC})_c^2 = {{ \pi n^2 }} }
The critical temperature is obtained by using the following
relationship between the field theory coupling $ g^2 _{NC} $ and the
the statistical mechanical coupling $ J $, namely,
\eqn\coupling{ {J \over T} = { 1 \over { g^2 _{NC}}}}
where $ J $ is the strength of the spin-spin interaction that occurs in
the expression for the spin-spin interaction energy $ E $ as follows
\eqn\energy{  E = J \sum_{ij} \cos (\theta_i - \theta_j) }
where $ \theta_i $ refers to the order parameter at position $ i $  in
a two dimensional lattice and the pair $ {ij} $ refers to nearest neighbours.
With the correspondence {\coupling } the critical temperature is
obtained as 
\eqn\critical{ T_c = {{ J \pi n^2 }  }}

\newsec{Effect of finite lattice size on KT transition }

In this section we will modify our previous calculations and take the
continuum limit while keeping the lattice size fixed. This limit has been considered in {\Ambjorn }, { \ref\makeenko{ Y. Makeenko, JETP Lett. 72 : 393 - 399, 2000, Pisma Zh. Eksp. Teor. Fiz. 72 : 566 - 576, 2000; hep-th/0009028. }.  We shall use the
same $ N $ dimensional representation of $ \Gamma _{\mu} $ as in
section 2. We expand the
matrix model field $ \bar \phi $ as follows,
\eqn\twentyseven{ \bar \phi = \sum_{n_x , n_y =  0 }^{{(N - 1) / \beta}}
{\phi _{n_x n_y}} {\Gamma _1^{\beta n_y}} {\Gamma _2^{\beta n_x}}
e^{-{i \over 2} n_x n_y \alpha \beta^2} \equiv \sum _{k_x k_y} {\tilde
\phi}(k_x,k_y) e^{-ik_{\mu}X^{\mu}}     } 
where $ \phi_{n_x n_y} = {\tilde \phi}(k_x,k_y) $ and  $ \beta \equiv { N \over N_0 } $ is an integer and $ N_0 $ is
the number of lattice sites along one dimension in the lattice. The upper limit on the
sum above follows from the fact that $ {\Gamma _ \mu}^ N = 1. $    For a lattice of spacing
$ a $ and size $ L = N_0 a, $ the dimensional momenta $ k_x $ and $
k_y $ are again given as in {\bsix}. The fourier expansion of the field $ \phi $
in dimensionless and dimensional coordinates is then as follows 
\eqn\fity{ { {\phi}}(\tilde x, \tilde y) = \sum_{n_x ,n_y } { \phi
_{n_{x} n_{y}}} e^{-i\alpha \beta ({n_x} \tilde x + {n_y}\tilde y)} =
\phi(x,y) =
\sum_{k_x , k_y } {
{\tilde \phi} ({k_x  k_y})} e^{-i(k_{x}x + k_{y}y)}}
The product of two
matrix model operators $ {\bar \phi} _1 $ and $ {\bar \phi} _2 $ is given by  
\eqn\sfiftyone{ \bar \phi_{1} \bar \phi_{2} =\sum_{{n_i}_x ,{n_i}_y =
0 }^{ (N - 1)/ \beta}{\phi_{{n_1}_x{n_1}_y}}
{\phi_{{n_2}_x{n_2}_y} } {\Gamma _1^{\beta  N_y }}{\Gamma _2^{ \beta N_x}} e^{-{i \beta^2 \alpha
\over 2} N_x N_y  } e^{-{i \beta^2 \alpha \over 2} ({n_1}_x {n_2}_y
- {n_2}_x {n_1}_y )}}
where
\eqn\twohundredeight{ N_x = ( n_{1x} + n_{2x} ), ~~~ N_y = (n_{1y} +
n_{2y}) }
Recalling the map from the reduced model to the noncommutative field
theory we obtain the following equation for $ \theta $
\eqn\hundred{ \theta = { N a^2 \over {2 \pi }}} 
The continuum limit ( $ a \rightarrow 0 $ ) is taken such that the
system size $ L = N_0 a $ and
the dimensional noncommutativity parameter $ \theta $ are kept
constant. This implies the following scaling relations
\eqn\hundredone{ a = {\sqrt {{ 2 \pi \theta } \over N }} ,~~~~ \beta =
{ 1 \over L }{\sqrt { 2 \pi \theta N }}}
In this case, the following relation holds between the integrals of
fields and traces of corresponding operators,
\eqn\hundredtwo{ Tr{\bar \phi } = N { \phi _{00}} = { N \over {{N_0}^2} } \sum _
{\tilde x, \tilde y} \phi (\tilde x , \tilde y ) = { N \over L^2 }
\int dx dy \phi (x,y)}
We now consider the noncommutative XY model action given by {\aone
}. As usual, the corresponding reduced model is obtained by replacing
derivatives by commutators, star products by matrix products and
integrals by traces. Noticing that the map between derivatives and
commutators remains the same as in {\bnine }, we get the following form
for the reduced action
\eqn\hundredthree{ S = - { 1 \over g^2 } Tr [P_ \mu , U^+] [P^ \mu , U]}      
where $ g^2 = {{N \over {L^2}} g^2_{NC} }.  $ An evaluation of the action
along lines analogous to sections (3) and (4) then gives 
\eqn\hundredfour{ S =  {2B \over {g^2}}[2\sum_{m=0}^{ L^2 / \theta }{\sqrt{m(m+n)}}
- \sum_{m=0}^{L^2 / \theta }(2m+1)]}
The upper limit on the sum follows from the fact that the
spreads of the harmonic oscillator levels $ |m> $ ( which scale as 
$ \sqrt {m \theta} $ ) should be bounded by the size  $
L $ of the system. Evaluating the sum in {\hundredfour } we obtain ( after
subtracting the spurious linear dependence on $ L^2 /\theta $), for large $ L^2/\theta $,
the following expression for the action,
\eqn\hundredfive{ S = - { B n^2 \over {2 g^2} } \ln {L^2 \over \theta
}} 
The entropy is given by the same expression as in the infinite lattice
case. The dimensionless Helmholtz free energy is, therefore, given by
\eqn\hundredseven{ F = -{  B n^2 \over {2g^2} } \ln {L^2 \over
\theta} - \ln {L^2 \over \theta } = -{  B n^2 L^2 \over {2 N g^2_{NC}} } \ln {L^2 \over
\theta} - \ln {L^2 \over \theta }      }
from which it follows that the critical values $ g_c $ and $ (g_{NC})_c $
of the couplings above which it is favourable to produce vortices is given by
\eqn\vort{ g^2_c = -{{B n^2 } \over 2 }, ~~~~ (g_{NC})^2_c = - {{B n^2
L^2} \over {2 N}   }}      
From {\twentyfive} and  {\vort} it is clear that the critical coupling
$ g_c $ is the same in both the finite and infinite lattice
cases. However, the crtitical coupling $ g_{NC} $, in the large $ N $
limit, is finite in the infinite lattice case and zero in the case of the
finite lattice. It, therefore, follows that the transition to the
vortex phase occurs at finite temperature for infinite lattice and at
zero temperature for the finite lattice.

\newsec{Conclusion}

The noncommutative $ X-Y $ model we have considered is a
generalization of the commutative $ X-Y $ model for nonzero
noncommutativity parameter $ \theta $. By mapping the field theory to its equivalent
matrix model we have been able to construct vortex solutions in
it. Kosterlitz Thouless transition from a phase of no vortex to a
phase of vortices occurs when the field theory coupling or,
equivalently, the matrix model coupling, reaches a critical value. We have
shown that the critical matrix model coupling is the same when the
system is put in either a finite or an infinite lattice. The critical
values of the coupling in the continuum field theory is finite in the
case of an infinite lattice and zero in the case of a finite lattice.

\newsec{Acknowlegdement}

I would like to thank S.R. Das for suggesting this problem to me and
for providing invaluable insights at every stage of the analysis.

\listrefs 
\end